\documentclass[aps,floats,floatfix,twocolumn,prb,showpacs,10pt,superscriptaddress]{revtex4-1}
\usepackage[T1]{fontenc}
\usepackage{graphicx}
\usepackage{color}
\usepackage{upgreek}
\usepackage{setspace}
\usepackage{amssymb}
\usepackage{amsmath}
\usepackage{pstricks}
\usepackage{dsfont}
\usepackage{fancyhdr}
\usepackage{array}
\usepackage{multirow}
\usepackage{booktabs}
\usepackage{diagbox}
\usepackage{hhline}
\usepackage{oubraces}
\usepackage{footmisc}
\usepackage{float}
\usepackage{soul}
\usepackage{hyperref}
\usepackage{verbatim}

\usepackage[english]{babel}

\allowdisplaybreaks[4]

\newcolumntype{M}[1]{>{\centering\arraybackslash}m{#1}}

\begin{document}
\title{Fluctuating local field method for the disordered Ising model}

\author{Daria Kuznetsova}
\affiliation{Russian Quantum Center, Skolkovo innovation city, 121205 Moscow, Russia}
\affiliation{Moscow Institute of Physics and Technology, Institutskiy per. 9, Dolgoprudny, Moscow Region, 141701, Russia}

\author{Grigory V. Astretsov}
\affiliation{Russian Quantum Center, Skolkovo innovation city, 121205 Moscow, Russia}
\affiliation{Department of Physics, Lomonosov Moscow State University, Leninskie gory 1, 119991 Moscow, Russia}

\author{Alexey N. Rubtsov}
\affiliation{Russian Quantum Center, Skolkovo innovation city, 121205 Moscow, Russia}
\affiliation{Department of Physics, Lomonosov Moscow State University, Leninskie gory 1, 119991 Moscow, Russia}
\date{\today}

\begin{abstract}
We present a method for computing thermal properties of classical spin clusters with arbitrarily chosen interactions between spins. For such systems, instability channels are \textit{a priori} not known. The method is based on the Fluctuating Local Field (FLF) approach, the effective description of the system with a nonlinear and fluctuating field, and achieves substantial improvements over mean-field theory. We show that two fluctuating modes are sufficient for numerically accurate solution of systems consisting of two dozen spins, while for larger systems it is needed to account for a larger number of fluctuating modes for a full quantitative agreement with the exact solution.
\end{abstract}

\maketitle

\section{Introduction}
\label{sec:intro}

It is well understood from the time of V. Ginzburg and L. Kadanoff that fluctuations of the order parameter are at heart of phase transformations and should be accounted in any theoretical description of these phenomena \cite{landau2013,wilson1974,kadanoff1966}. Significant progress has been achieved in this area,  mostly with various versions of the renormalization group technique \cite{wilson1983}, including recent advances with the functional renormalization group \cite{metzner2012,Kugler2018}. 

However, quantitative microscopic description of systems with competing order parameters are still complicated. Modelling of such systems as cuprate-based supercunductors  \cite{bednorz1986,dong2008}, and  frustrated magnetism \cite{lacroix2011} remains a great challenge till now. Low-dimension and nano-scale systems are particularly difficult in this context as they exhibit strong fluctuations of the order parameter(s) \cite{hermann2017,schaefer2021}.


Most of the methods use the mean field treatment   as a starting point and then consider fluctuations using perturbation theory \cite{rohringer2018}. One of the challenges to be mentioned is that the leading channel of fluctuations is not always known. One should construct an unbiased approach. For correlated fermionic systems, parquet equations \cite{Bickers2004} and functional renormalization group \cite{metzner2012} are used to treat fluctuations in different channels on the same footing. However, one of the difficulties of such unbiased diagrammatic treatment of fluctuations is its high computational complexity. For example, the parquet equations can be numerically solved for a cluster limited to $6\times6$ sites \cite{wentzell2020}, which is comparable to the performance of numerically exact methods limited by the exponential growth of the Hilbert space \cite{liu2021}. Significant increase of the performance requires additional approximations, such as integration out of high frequencies in the dual-fermion space \cite{astretsov2020}, optimized momentum grids \cite{eckhardt2020} or partial bosonization of vertex functions \cite{krien2020}. An additional problem is related to the fact that both parquet equations and functional renormalization group rely on the assumptions that collective fluctuations obey Gaussian statistics and/or their magnitude is small. Contrary,  small and low-dimensional systems can develop strong nonlinear collective modes -- antiferromagnetism in the Hubbard model at half filling is a good example here. Thus, use of the unbiased diagrammatic treatment of fluctuations for small and low-dimensional systems remains limited.


The Fluctuating Local Field (FLF) method \cite{rubtsov2018} is an alternative approach aiming a description of fluctuations regardless of their magnitude and statistics. In this approach, nonlinear  fluctuations of the soft mode(s) are described by introducing artificial classical fields acting on the corresponding degrees of freedom. First, the method has been formulated for a quantitatively good description of small classical lattices subjected to an external polarizing field. Later, it was extended to fermionic systems and applied to Hubbard-like clusters in 1D and 2D geometries \cite{rubtsov2020,lyakhova2021,linner2022}.  

So far, the Fluctuating Local Field approach has been formulated and applied to the systems where the leading channel of fluctuations is known {\it a priori}. This means that one had to find the leading modes by other methods and only afterwards use the FLF approach to accurately treat them. As we discussed, it would be desirable to construct an unbiased version of the FLF, so that the fluctuating modes are detected and treated within the same procedure. This is the goal of the present work.

In this paper, we introduce a self-consistent procedure to find leading instability channels within the FLF approach without resorting to other methods. We apply our approach  to  perform unbiased FLF calculations for Ising-class systems, not limited to certain geometries and particular types of interactions. We show that this approach allows to capture physics of classical spin clusters with a frozen disorder using only a few fluctuating modes.

\section{Model and method}
\label{sec:formalism}

We consider statistical properties of the classical Ising model with arbitrarily chosen interactions $J_{ij}$ between $N$ spins in the presence of a small magnetic field $h_i$:
\begin{equation}\label{eq:ising_energy}
    E =  - \sum_{i,j = 1}^N J_{ij} s_i s_j - \sum_{i = 1}^N h_i s_i, 
\end{equation}
where $E$ is the energy of a configuration of $N$ spins $\{s_i=\pm 1\}$, $i = 1,\dots N$. 
We will consider several systems with different interaction matrices. In particular, disordered models will be studied where leading channels of fluctuations are not {\it a priori} known. The on-site magnetic field $h_i$ is sampled from the Gaussian distribution with zero mean and a variance several orders of magnitude smaller than the interaction strength between spins. The purpose of the magnetic field is to lift possible degeneracies, so that in the limit of $T\rightarrow 0$ thermodynamic averages $\langle s_i \rangle$ are either $+1$ or $-1$. For higher temperature, a response to $h$ allows to guess about the linear susceptibility of the system.

Let us present the description of the Fluctuating Local Field method for finding finite temperature properties of the model under study. We will formulate the FLF method using an auxiliary problem which describes an effective long-range interaction between spins that comes from low-energy excitations of the system. The auxiliary problem is characterized by the energy 
\begin{equation}\label{eq:flf_energy}
	\tilde{E}= - \left(\sum_{i = 1}^N f_i s_i\right)^2  - \sum_{i = 1}^N h_i s_i,
\end{equation}
where $f_i$ are real-valued variational parameters which are to be self-consistently defined later. 
 
The system described by (\ref{eq:flf_energy}) can be readily solved by applying the Hubbard-Stratonovich transformation to the partition function $\tilde{Z}$ of the auxiliary problem at temperature $T$:
\begin{equation} \label{eq:hubbard_stratonovich}
    \begin{split}
        \tilde{Z} = \sum_{\{s\}} e^{- \tilde{E}/T} = \sum_{\{s\}} e^{\left(\left(\sum f_i s_i\right)^2 +\sum h_i s_i\right)/T} &= \\
	       \frac{1}{\sqrt{\pi T}}  \int dx\, e^{\left(-x^2 + 2 x \sum f_i s_i +\sum h_i s_i\right)/T} &= \int dx \,\tilde{z}(x),
	\end{split}
\end{equation}
where the integration goes from $-\infty$ to $\infty$, and
$$\tilde{z}(x)= \frac{1}{\sqrt{\pi T}} e^{-x^2/T} \prod_{i = 1}^N 2 \cosh \frac{2f_i x + h_i}{T}.$$ Similarly, one can also calculate the local magnetization $$\langle s_i \rangle = \frac{1}{\tilde{Z}}\int dx\, \tanh\frac{2f_i x + h_i}{T} \tilde{z}(x)$$ and correlation functions between the different sites $$\langle s_i s_j \rangle = \frac{1}{\tilde{Z}}\int dx\, \tanh\frac{2f_i x + h_i}{T} \tanh\frac{2f_j x + h_j}{T} \tilde{z}(x).$$
	
Variational parameters $f_i$ can be found using the self-consistency condition, which we  derive from the Gibbs-Bogolyubov-Feynman variational principle \cite{feynman1998}: 
\begin{equation}\label{eq:feynman}
    \langle E - \tilde{E} \rangle _{\tilde{E}} - T\log \tilde{Z} = \text{min}.
\end{equation}
Here $E$ is the energy of the original problem (\ref{eq:ising_energy}), $\tilde{E}$ is the energy of the trial problem (\ref{eq:flf_energy}) and averaging is performed with respect to the trial problem $ \langle \dots \rangle _{\tilde{E}} = \sum_{\{s\}} \dots e^{-\tilde{E}/T}/\tilde{Z}$. Thus, the self-consistency condition corresponds to a vanishing variation of (\ref{eq:feynman}):
\begin{equation}\label{eq:self-consistency}
	\sum_{i,j} (f_i f_j - J_{ij}) \delta \langle s_i s_j \rangle_{\tilde{E}} = 0.
\end{equation}

Solving equations (\ref{eq:self-consistency}) without any approximations is technically as hard as finding the global minimum of (\ref{eq:feynman}). However, if we assume that the largest contribution to correlators comes from the soft mode(s), this would significantly simplify the problem. In that case we are able to approximate the correlation function by its largest eigenvalue $\langle s_i s_j \rangle_{\tilde{E}} \approx \lambda\, v_i v_j$, where $\lambda$ is the maximum eigenvalue of $\langle s_i s_j \rangle_{\tilde{E}}$ and $v_i$ is the $i$-th component of the corresponding eigenvector $\mathbf{v}$. Therefore, we can rewrite $\delta \langle s_i s_j \rangle_{\tilde{E}}$ as
\begin{equation}
\begin{split}
    \delta \langle s_i s_j \rangle_{\tilde{E}} = \delta \left(\sqrt{\lambda}\, v_i \, \sqrt{\lambda}\, v_j\right)=\\ \delta\left(\sqrt{\lambda}\, v_i \right) \sqrt{\lambda}\, v_j + \sqrt{\lambda}\, v_i \, \delta\left(\sqrt{\lambda}\, v_j \right).
\end{split}
\end{equation}
This brings us to the self-consistency condition
\begin{equation}\label{eq:self-consistency-2}
	\sum_{j = 1}^N (f_i f_j - J_{ij}) v_j = 0,
\end{equation}
which holds for any $i$.

The set of $N$ equations (\ref{eq:self-consistency-2}) is more feasible than the condition (\ref{eq:self-consistency}), and  requires only a calculation of the leading  eigenvector  $\mathbf{v}$ of the second-order correlator matrix $\langle s_i s_j \rangle$.

One way to obtain $f_i$ satisfying the obtained self-consistency condition is to solve (\ref{eq:self-consistency-2}) iteratively:
\begin{equation}\label{eq:iter}
	f_i ^{(n+1)} = \frac{\sum_{j} J_{ij} v^{(n)}_j}{\sum_{j} f_j ^{(n)} v^{(n)}_j},
\end{equation}
where $(n)$ is the iteration number. In our numerical implementation of the method, we have overcome possible instabilities arising in (\ref{eq:iter}) by introducing damping factors $\alpha$ for updates $f_i^{(n)} = (1-\alpha)f_i^{(n)} + \alpha f_i^{(n-1)}$. For the most unstable regimes of parameters we mixed all previous iterations together $f_i^{(n)} = \sum_{m=1}^n \alpha^{(m)} f_i^{(m)}$, provided $\sum_{m=1}^{n} \alpha^{(m)}=1$. The latter scheme is guaranteed to have a stable fixed point, but the rate of convergence may be slower. During the computations, we started from the paramagnetic phase and then gradually decreased temperature. Variational parameters which were obtained for a higher temperature were used as an initial guess for solving self-consistency equations for a lower temperature.
	
We can extend the method and take into account several low-energy modes. For the case of two modes the total two-mode FLF energy is
\begin{equation}\label{eq:FLF2}
	\tilde{E}_2 =  - \left(\sum_{i = 1}^N f'_i s_i\right)^2 - \left(\sum_{i = 1}^N f''_i s_i\right)^2 -  \sum_{i = 1}^N h_i s_i,
\end{equation}
where the adjustable parameters are now the two-component vectors  ${\bf f}_i=(f'_i, f''_i)$.

Hubbard-Stratonovich decoupling yelds the trial partition function:
\begin{equation}\label{eq:z2}
	    \tilde{Z}_2 = \int d^2x\, \tilde{z}_2 ({\bf x});
\end{equation}
$$\tilde{z}_2({\bf x})= \frac{1} {\pi T} e^{-\mathbf{x}^2/T} \prod_{i = 1}^N 2 \cosh\frac {2 \mathbf{f}_i \mathbf{x} + h_i}{T},$$ where $\mathbf{f}_i \mathbf{x}= f'_i x'+f''_i x''$ is the scalar product.
The expressions for averages stay formally the same as for the single-mode FLF (see Eqs. (\ref{eq:hubbard_stratonovich})), with the only substitution of $f_i x$ for $\mathbf{f}_i \mathbf{x}$. The self-consistency condition then reads
\begin{equation}\label{eq:mean2}
	\sum_{j = 1}^N (\mathbf{f}_i \mathbf{f}_j - J_{ij}) \delta \langle s_i s_j \rangle_{\tilde{E}_2} = 0,
\end{equation}
where $\mathbf{f}_i \mathbf{f}_j = f'_i f'_j + f''_i f''_j$. 

Now assume that the strongest contribution to the correlation function comes from two lowest modes $\langle s_i s_j \rangle \approx \lambda^{(1)}  v^{(1)}_i v^{(1)}_j + \lambda^{(2)} v^{(2)}_i v^{(2)}_j $. The simplified version of (\ref{eq:mean2}) reads
\begin{equation}\label{eq:self_consistency_two-mode-flf}
	\sum_{j} (\mathbf{f}_i \mathbf{f}_j - J_{ij}) v^{(1,2)}_j = 0,
\end{equation}
which is now a set of $2N$ equations for $2N$ variables. 
The generalization of (\ref{eq:z2}) and (\ref{eq:self_consistency_two-mode-flf}) for an arbitrary number low-energy modes is straightforward.

\section{Results}

We applied one-mode and two-mode Fluctuating Local Field method to the Ising clusters of sizes ranging from 16 to 32 sites with different types of interaction between spins, and a small external field $h_i$ sampled from a Gaussian distribution with a variance of $0.03$ for all of the discussed problems. We calculated response functions to the magnetic field and compared them to the response functions calculated by means of exact enumeration (ED) and the mean-field theory (MF). As part of the necessary checks, we made sure that both FLF and MF methods give the true ground state energy for all systems examined. 

The response of a local magnetization $\langle s_i \rangle$ to an external non-uniform magnetic field $h_j$ is described by the the susceptibility matrix 
\begin{equation}
    \chi_{ij} = T \frac{\partial \langle s_i \rangle} {\partial h_j}.
\end{equation}
In the thermodynamic limit, when a system experiences a phase transition, the leading eigenvalue $\Lambda_\text{max}$ of the matrix $\chi_{ij}$ diverges. Finite systems do not display phase transition and eigenvalues of $\chi_{ij}$ are limited to the size of the system, $N$. It is worth pointing out that MF predicts phase transition, and hence divergent response functions, even for finite-size systems. The mean-field susceptibility $\chi^\text{mf}_{ij}$  is given by the matrix
\begin{equation}
\chi^\text{mf}_{ij} = \left (\frac{1}{1 - \langle s_i \rangle^2} \delta_{ij} - \frac{2}{T} J_{ij} \right)^{-1},
\end{equation}
where $\delta_{ij}$ is the Kronecker delta. This matrix always has the infinite eigenvalue at the temperature $T^\text{mf}_c = 2 \mathcal{J}$, where $\mathcal{J}$ is the largest eigenvalue of the coupling matrix $J_{ij}$.

\begin{figure}[t!] 
    \begin{center}
        \includegraphics[width=\linewidth]{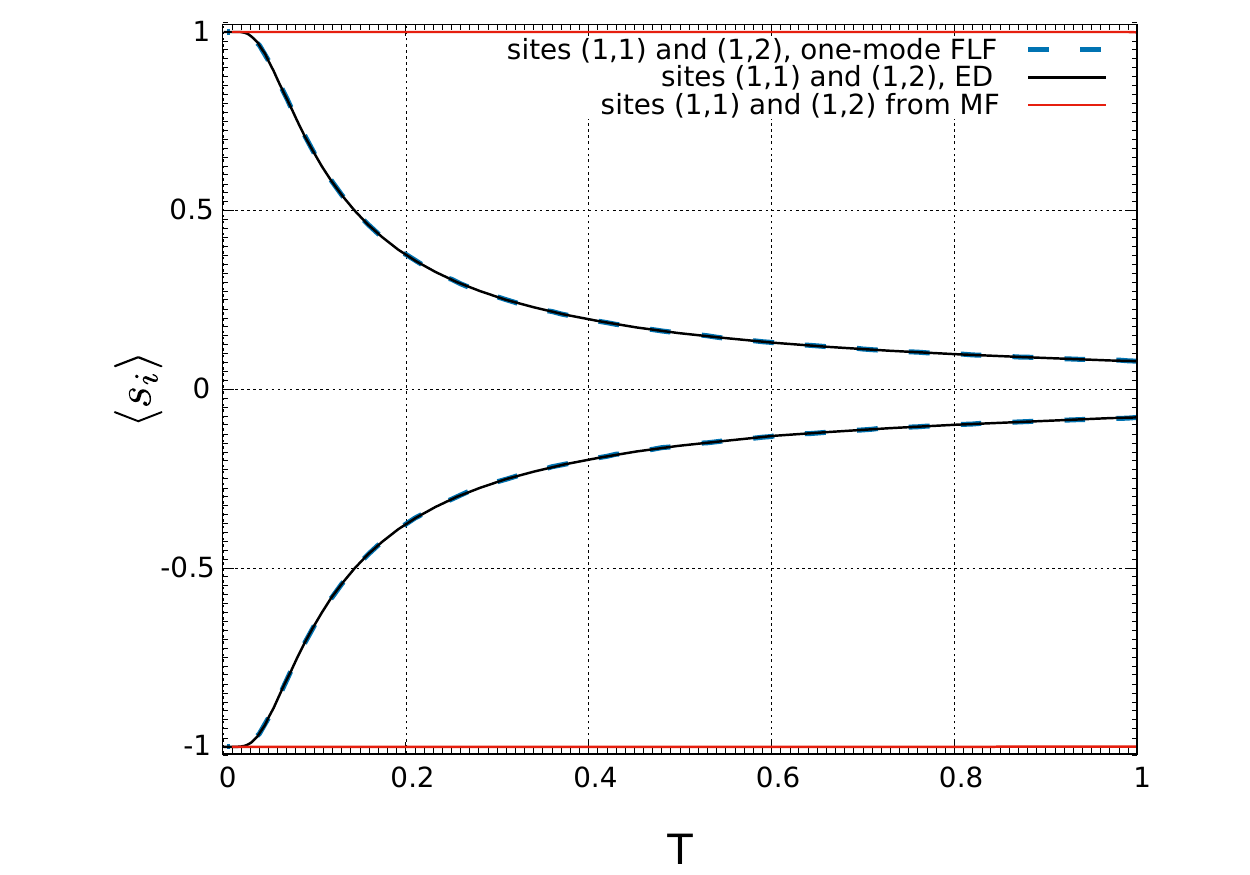}
        \caption{Dependence of on-site magnetization on temperature for two neighboring spins in a $4\times 4$ antiferromagnetic cluster with periodic boundary conditions obtained by one-mode FLF (blue dashed), MF (red) and exact (black) methods.}
        \label{Fig:16-square-magn}
    \end{center}
\end{figure}

\begin{figure}[t!] 
    \begin{center}
        \includegraphics[width=\linewidth]{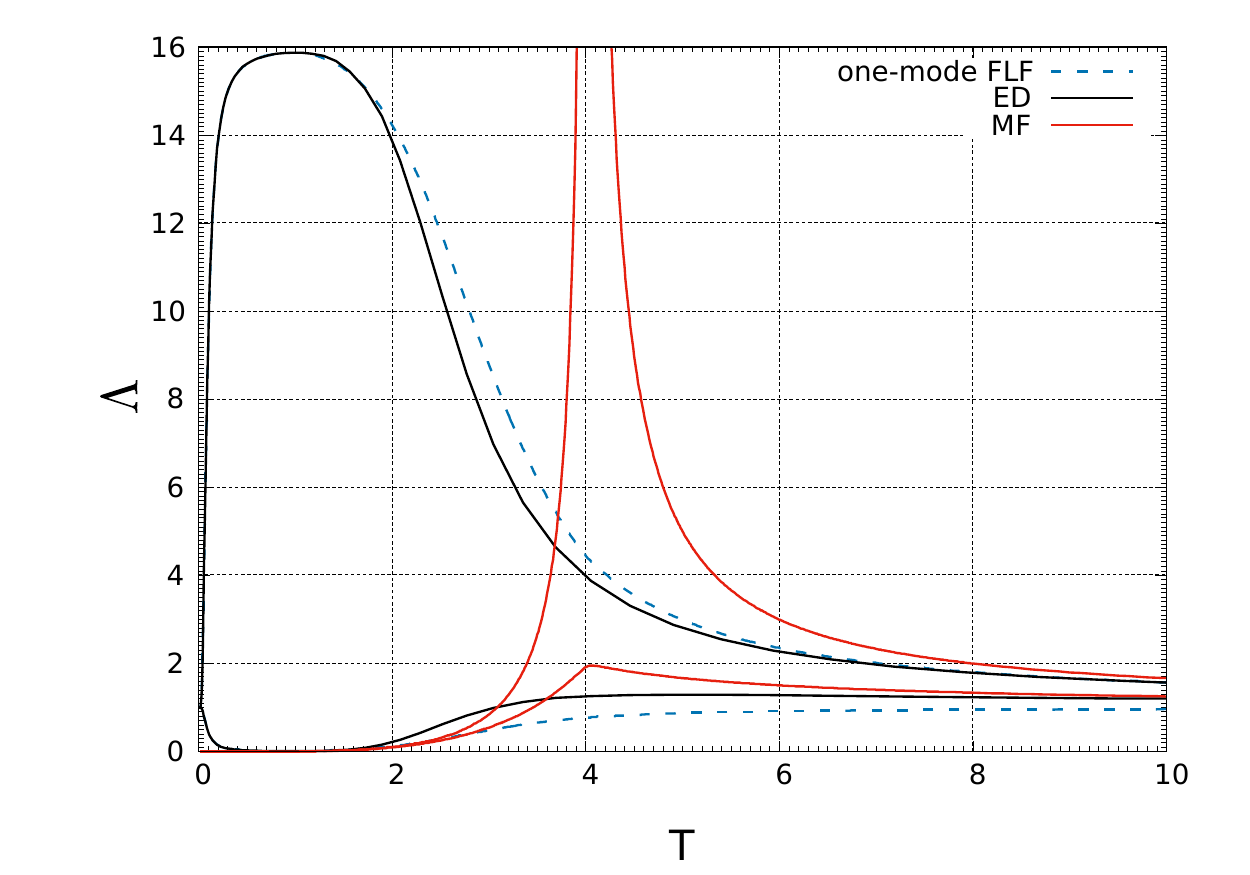}
        \caption{Dependence of two leading eigenvalues of magnetic susceptibility on temperature in a $4\times 4$ antiferromagnetic cluster with periodic boundary conditions obtained by one-mode FLF (blue dashed), mean field (red) and exact (black) methods.}
        \label{Fig:16-square-ev}
    \end{center}
\end{figure}

We start our analysis by considering a $4\times 4$ cluster with uniform antiferromagnetic couplings between neighboring spins ($J_{\langle i,j \rangle}=-1$) subjected to periodic boundary conditions. This system shows a checkerboard pattern of spins at zero temperature. Figure \ref{Fig:16-square-magn} shows a temperature dependence of local spin polarization for neighboring sites whose $(x, y)$ coordinates are $(1,1)$ and $(1,2)$. We notice a perfect agreement between the one-mode FLF method and the exact solution. One naturally expects that for small clusters the mean-field approximation gives overestimated values of the order parameter. In this case, MF predicts an already saturated local magnetization, while in fact it is set only at sufficiently low temperatures. Figure \ref{Fig:16-square-ev} shows the temperature dependence of two leading eigenvalues $\Lambda$ of the magnetic susceptibility matrix $\chi_{ij}$ One can see that a single fluctuating soft mode is enough to describe physics of this system. Slight deviations in the behaviour of the leading eigenvalue in one-mode FLF from the exact solution are visible in the range of temperatures from 2 to 6. The second largest eigenvalue for FLF qualitatively repeats the behavior of the exact one but its value is somewhat lower. This is to be expected since the simplified self-consistency (\ref{eq:self-consistency-2}) was derived under the assumption that the largest contribution to fluctuations comes from the biggest eigenvalue. One can also notice that mean-field eigenvalues are quite close to the exact ones in the region of high temperatures, but then erroneously predict a phase transition at $T^\text{mf}_c=4$.

\begin{figure}[t!] 
    \begin{center}
        \includegraphics[width=\linewidth]{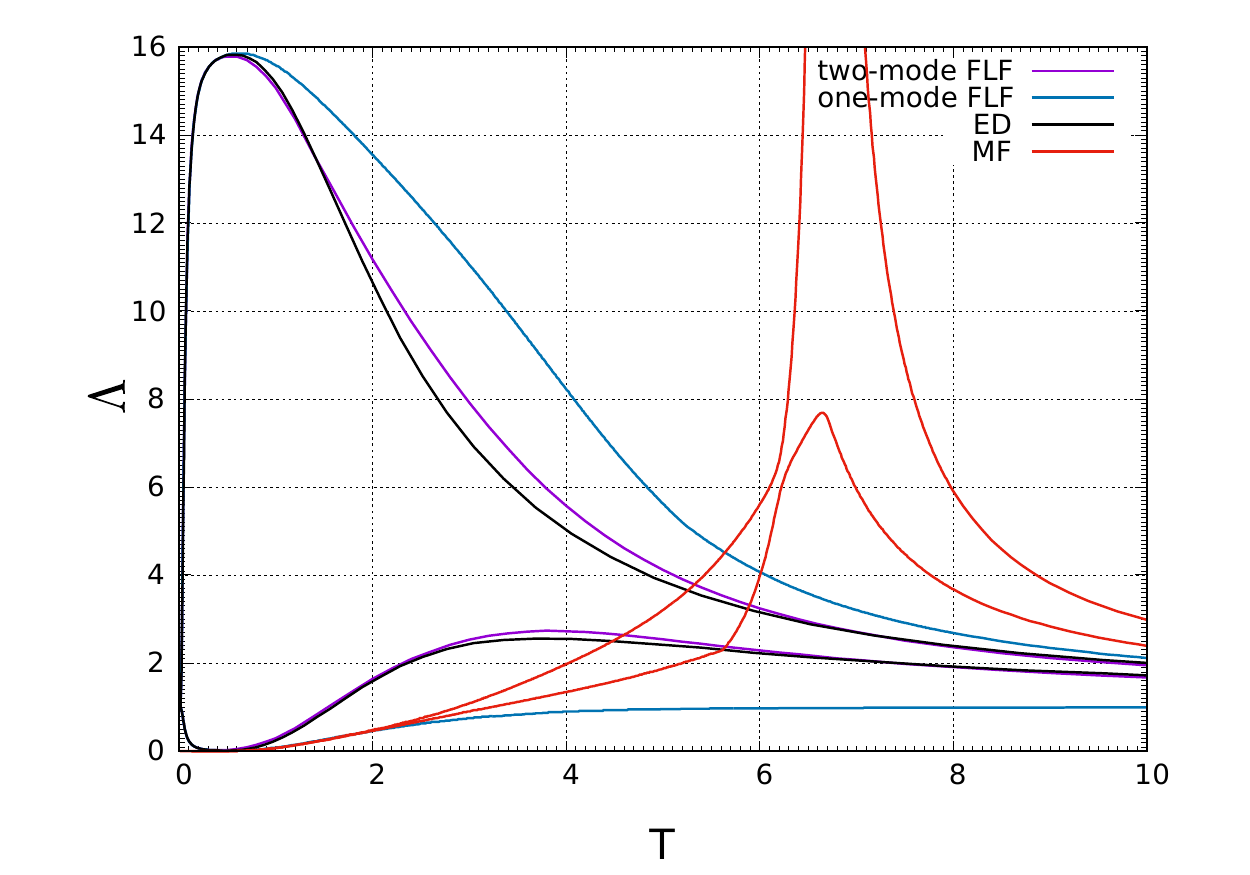}
        \caption{Dependence of two leading eigenvalues of magnetic susceptibility on temperature in a 16 sites cluster with antiferromagnetic binary random couplings between all spins obtained by two-mode FLF (purple), one-mode FLF (blue), mrean field (red) and exact (black) methods.}
        \label{Fig:eigenvals_16}
    \end{center}
\end{figure}

In order to illustrate how disorder affects the behavior of the susceptibility, we now consider a cluster formed by 16 spins with all-to-all random binary antiferromagnetic interactions, i.e. values of $J_{ij}$ are either $-1$ or $0$ with a probability 50\%. Fig. \ref{Fig:eigenvals_16} shows the temperature behavior of two largest eigenvalues of spin susceptibility for this model. We observe that this time one fluctuating mode only qualitatively describes the behavior of the system. It correctly predicts the point where fluctuating are the strongest but overestimates their magnitude for higher temperatures. Introducing the second mode helps resolve quantitative disagreements between FLF and the exact solution. We note that for the two-mode FLF not only the largest eigenvalue, but also the second largest one is close to the exact solution.

\begin{figure}[t!] 
    \begin{center}
        \includegraphics[width=\linewidth]{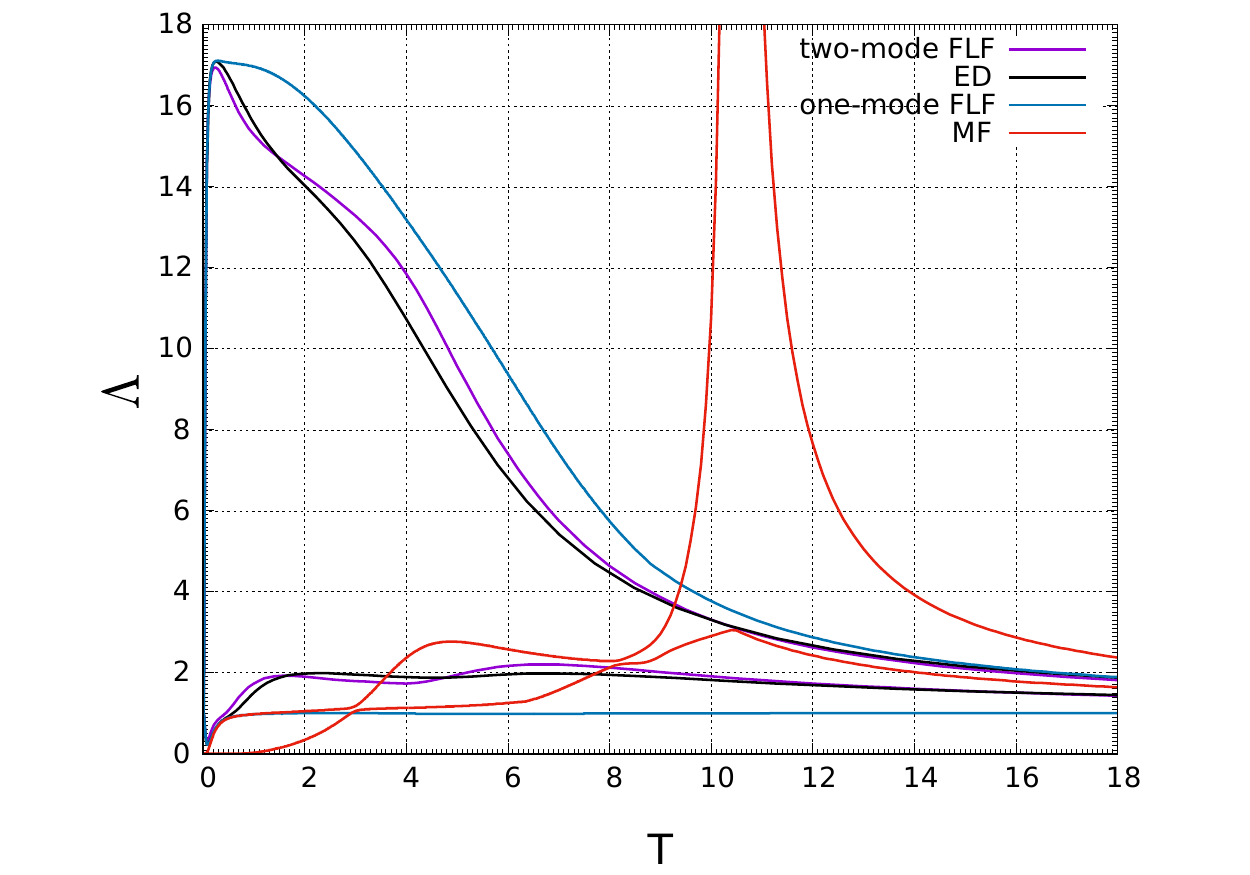}
        \caption{Dependence of two leading eigenvalues of magnetic susceptibility on temperature in a 18 sites cluster with Gaussian random couplings between all spins obtained by two-mode FLF (purple), one-mode FLF (blue), mean field (red) and exact (black) methods.}
        \label{Fig:eigenvals_18}
    \end{center}
\end{figure}

If we let the interaction matrix elements have an arbitrary sign, the shape of the temperature dependence of the magnetic response will be more complicated. Fig. \ref{Fig:eigenvals_18} shows eigenvalues of $\chi_{ij}$ for a cluster consisting of 18 sites with random interactions between all spins sampled for the Gaussian Orthogonal Ensemble with zero mean and variance 1. One can see that one-mode FLF once again correctly predicts the temperature where fluctuations are the strongest while failing to reproduce features in the shape of the temperature dependence of leading eigenvalues. However, the two-mode FLF successfully captures them for both eigenvalues.

\begin{figure}[t!] 
    \begin{center}
        \includegraphics[width=\linewidth]{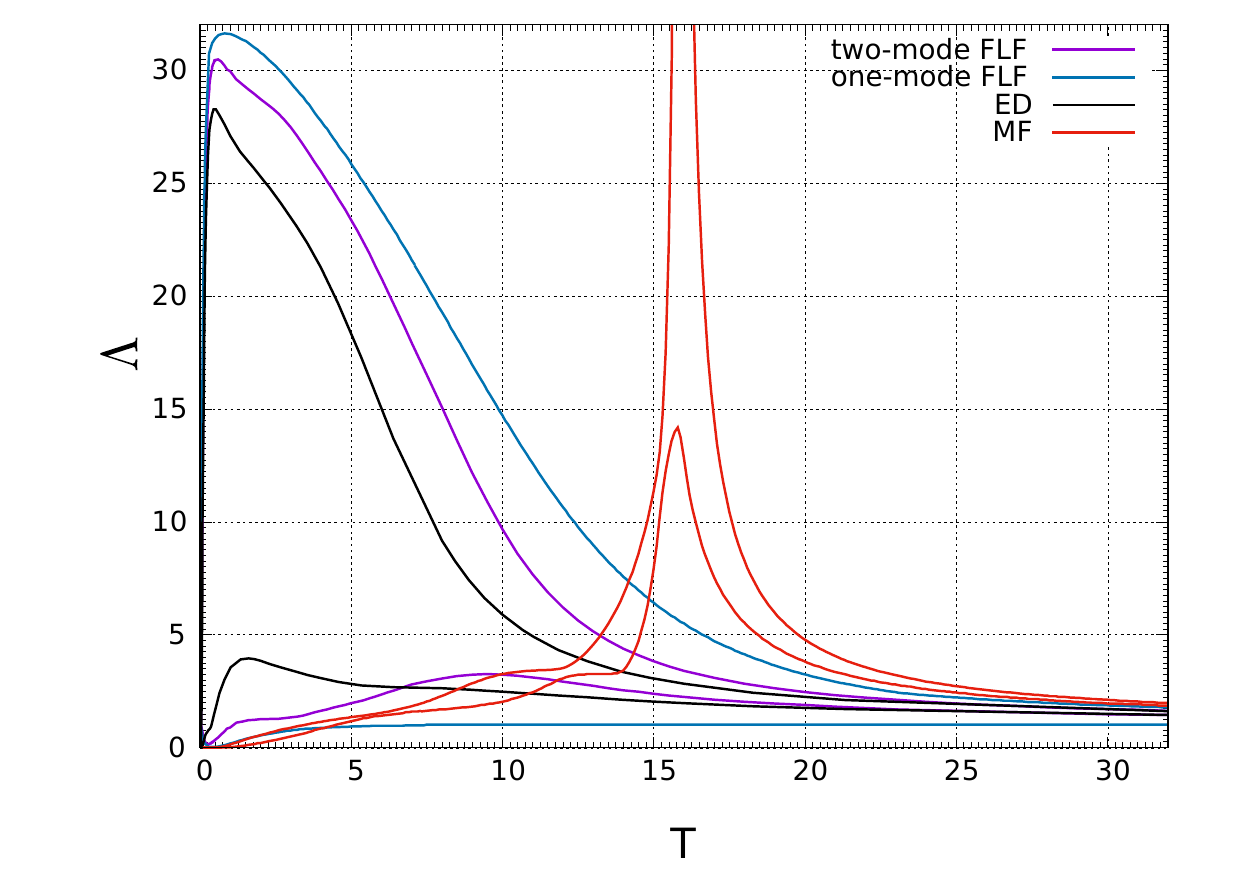}
        \caption{Dependence of two leading eigenvalues of magnetic susceptibility on temperature in a 32 sites cluster with Gaussian random couplings between all spins obtained by two-mode FLF (purple), one-mode FLF (blue), mean field (red) and exact (black) methods.}
        \label{Fig:eigenvals_32}
    \end{center}
\end{figure}

Fig. \ref{Fig:eigenvals_32} shows the temperature dependence of two leading eigenvalues of magnetic susceptibility for a cluster consisting of 32 spins with random interactions between all spins sampled for the Gaussian Orthogonal Ensemble with zero mean and variance 1. One-mode FLF strongly overestimates the role of fluctuations of a single mode. Two-mode FLF gives the leading eigenvalue somewhat closer to the exact solution yet failing to reproduce the behavior of the second largest eigenvalue of the response function. This suggests that it is necessary to include more than two fluctuating modes clusters of this size.

\section{Conclusions and Outlook}
\label{sec:conclusions}

In conclusion, we have developed the Fluctuating Local Field method to treat in an unbiased way classical spin systems with arbitrary interactions for which the instability channel is  \textit{a priori} not known. We showed that two low-energy modes is enough to capture the physics of clusters of 16 to 18 spins with local and all-to-all interactions, while clusters consisting of 32 spins require the incorporation of more modes for quantitatively accurate results. Therefore, the FLF method targets the class of systems whose size is relatively large and exact calculations are cumbersome. Extended to correlated fermionic systems, the method can applied to the study of complex molecules, atom cluster and SYK-type models \cite{chowdhury2022}.  

\paragraph*{Acknowledgements}
This research work was supported by the Roadmap for the Development of Quantum Technologies, contract No. 868-1.3-15/15-2021, dated October 5, 2021.

\bibliographystyle{apsrev4-1}
\bibliography{main-bib}

\end{document}